\documentclass[12pt]{article}  

\def\R{ {\rm R \kern -.31cm I \kern .15cm}}
\def\C{ {\rm C \kern -.15cm \vrule width.5pt \kern .12cm}}
\def\Z{ {\rm Z \kern -.27cm \angle \kern .02cm}}
\def\N{ {\rm N \kern -.26cm \vrule width.4pt \kern .10cm}}
\def\1{{\rm 1\mskip-4.5mu l} }
\def\lsim{\raise0.3ex\hbox{$<$\kern-0.75em\raise-1.1ex\hbox{$\sim$}}}
\def\gsim{\raise0.3ex\hbox{$>$\kern-0.75em\raise-1.1ex\hbox{$\sim$}}}
\def\noi{\noindent}

\def\beq{\begin{equation}}   \def\eeq{\end{equation}}
\def\bea{\begin{eqnarray}}  \def\eea{\end{eqnarray}}

\def\noi{\noindent}
\def\beeq{\begin{eqnarray}} \def\eeeq{\end{eqnarray}}

\begin{document}

\begin{center} {\large \bf The Absence of Positive Energy Bound States } \vskip 3 truemm

 {\large \bf for a Class of Nonlocal Potentials} \vspace{1 truecm}

{\bf Khosrow Chadan}\\ {\it Laboratoire de Physique
Th\'eorique}\footnote{Unit\'e Mixte de Recherche UMR 8627 - CNRS }\\   
{\it Universit\'e de Paris XI, B\^atiment 210, 91405 Orsay Cedex,
France} \par \vskip 8 truemm

{\bf and} \par \vskip 8 truemm

{\bf Reido Kobayashi} \\ {\it
Department of Mathematics}\\ {\it  Tokyo University of Science, Noda,
Chiba 278-8510, Japan} \par \vskip 2 truemm
\end{center}

\begin{abstract}
We generalize in this paper a theorem of Titchmarsh for the
positivity of Fourier sine integrals. We apply then the theorem to
derive simple conditions for the absence of positive energy bound
states (bound states embedded in the continuum) for the radial
Schr\"odinger equation with nonlocal potentials which are superposition
of a local potential and separable potentials.
\end{abstract}
\vskip 2 truecm

\centerline{\it D\'edi\'e \`a Michel Gourdin et Andr\'e Martin pour leurs soixante-quinze ans.}
\vskip 3 truecm
\begin{flushleft}
LPT Orsay 04-67 \\
Ao\^ut 2004
\end{flushleft}
 
\newpage
\pagestyle{plain}
\baselineskip 20pt
\noi {\bf I. Introduction} \\

 Nonlocal separable two-body interactions have often been used in
nuclear physics and many-body problems because of the fact that the
two-body Schr\"odinger equation is easily solvable for them, and leads
to closed expressions for a large class of such interactions. They have
also been used very systematically with Faddeev equations for the
three-body problem. Their main feature is that the partial-wave $t$-matrix
has a very simple form, and can be continued off the energy-shell in a
straightforward manner, a feature which is most important, as is well
known, in nuclear physics, and in the Faddeev equations.$^1$ The only
problem with such potentials is the existence of positive energy bound
states, i.e. bound states embedded in the continuous spectrum.$^{2,3}$ This
is a general feature with nonlocal potentials, whether short-range or
not, contrary to the case of local potentials for which positive energy
bound states exist only if the potential is long-range and oscillating
at infinity.$^{1,4}$ Such states are, of course, undesirable, and
should be avoided. Their main feature is that they are highly unstable,
in the sense that a slight change in the potential makes them
disappear, or shifts them far away, whereas, for usual bound states
with negative energy, i.e. below the continuous spectrum, we have the
continuity theorem.$^{1,4}$ \par

Although nonlocal separable potentials have been used for decades now,
as said earlier, the only paper we know in which the absence of
positive energy bound states is shown for a particular class of
nonlocal potentials is the paper of Zirilli,$^5$ in which the author
shows the absence of such states for general nonlocal potentials which
are dilatation analytic in the sense of Combes. The purpose of the
present paper is to give other simple conditions for the absence of
these states for nonlocal potentials which are the sum of a local
potential and a separable potential.\par

The three-dimensional Schr\"odinger equation for the scattering of a
particle by a general nonlocal interactions reads

\beq \label{1e}  
(\Delta + E) \Psi (\vec{k}, \vec{r}) = \int U(\vec{r}, \vec{r}{\, '})\Psi (\vec{k}, \vec{r}{\, '}) d\vec{r}{\, '} \ .
\eeq

\noi Separable interactions are those for which 

\beq 
\label{2e}  
\left \{ \begin{array}{l} U(\vec{r}, \vec{r}{\, '}) = \displaystyle{\sum_{l=0}^{\infty}} \ \displaystyle{\sum_{n=1}^{N_l}} \varepsilon_{nl}\ u_{nl}(r) \ u_{nl}(r') \ P_l (\cos \theta ) \ , \\ \\ r = |\vec{r}|\ , \quad r' = |\vec{r}{\, '}|\ , \quad \cos \theta = \displaystyle{{\vec{r} \cdot \vec{r}{\, '} \over rr'}}\ , \quad \varepsilon_{nl} = \pm 1 \ . \end{array} \right .
\eeq

A more general class consists of separable interactions plus a local 
potential $V(r)$, which we assume to be spherically symmetric. \\

\noi \underline{\bf Remark.} As it is seen here, changing each $u$ to
$- u$ does not change the potential, and hence the equation. This is
the reason why one had to add also the $\varepsilon_{nl}$. It can be
seen that $\varepsilon = 1$ corresponds to a repulsive interaction,
whereas $\varepsilon = -1$ leads to an attractive one.$^{2,3}$\\

In the present paper, we shall consider the case where only one
separable term is present in each angular momentum state~:

\beq \label{3e} U(\vec{r}, \vec{r}{\, '}) = \sum_l \varepsilon_l \ u_l (r)\
u_l(r') \ P_l(\cos \theta ) + V(r) \ \delta (r - r')\ . \eeq

\noi It is for this class of potentials that we are going to obtain
simple conditions for the absence of positive energy bound states. \par

As usual, in order to secure the self-adjointness of the Hamiltonian,
and the existence of a decent scattering theory, one must impose some
conditions on $u_l(r)$ and $V(r)$. It turns out that sufficient
conditions for being on the safe side are the followings~:$^{2,3}$

\beq 
\label{4e}  
\left \{ \begin{array}{l} \hbox{$u_l(r)$ and $V(r)$ are both real, and locally $L^1$ for $r\not= 0$, $V(r) \geq 0$}\ , \\ \\ \displaystyle{\int_0^{\infty}} r^2|u_l(r)|dr < \infty \ , \quad \displaystyle{\int_0^{\infty}} r\ V(r) \ dr < \infty \ . \end{array} \right .
\eeq

Making use of the partial wave decomposition

\beq
\label{5e}
\Psi (\vec{k}, \vec{r}) = \sum_{l=0}^{\infty} (2 l + 1) i^l \ {\psi_l (k, r) \over kr}\ P_l (\cos \theta )\ ,
\eeq

\noi we obtain the radial Schr\"odinger equation

\beq 
\label{6e}  
\left \{ \begin{array}{l} \left [ \displaystyle{{d^2 \over dr^2}} + k^2 - \displaystyle{{l(l+1) \over r^2}}\right ] \psi_l (k, r) =  \varepsilon_{l}\ U_{l}(r) \displaystyle{\int_0^{\infty}} \\ \\ U_l(r') \ \psi_l(k, r')dr' + V(r) \ \psi_l (k, r)\ ,\\  \\ U_l (r) = (4 \pi )^{1/2} r\ u_l(r)\ , \quad \psi_l(k, 0) = 0\ . \end{array} \right .
\eeq
\vskip 5 truemm

For simplicity, we begin with the $S$-wave ($l = 0$). We shall see
later how to generalize the results to higher waves. Consider now first
the case where we have no local potential $V$ present~: 

\beq 
\label{7e}  
\left \{ \begin{array}{l} \psi '' + k^2 \psi = \varepsilon\ U(r)  \displaystyle{\int_0^{\infty}} \psi (k, r') U(r') dr' \ , \\ \\ 
\varepsilon = \pm 1 \ , \qquad \displaystyle{\int_0^{\infty}} r|U(r)|dr < \infty \ . \end{array} \right .
\eeq

\noi It can then be shown that the positive energy bound states with
energies $k_{\nu}^2$ ($k_{\nu} > 0$) are given by the simultaneous
roots of the following two equations$^{2,3}$

\beq 
\label{8e}  
\left \{ \begin{array}{l}  \widetilde{U}(k_{\nu}) = 0 \ , \\ \\ 
\varepsilon +  \displaystyle{{2 \over \pi}} \ P  \displaystyle{\int_0^{\infty}} \displaystyle{{\widetilde{U}^2(p) \over p^2 - k_{\nu}^2}}\ p^2 dp = 0\ , \end{array} \right .
\eeq

\noi where

\beq
\label{9e}
\widetilde{U}(p) = \int_0^{\infty} U(r) \ {\sin pr \over p}\ dr \ , 
\eeq

\noi and $P$ means the principal value of the integral. Under our
conditions (\ref{7e}) on $U(r) = \sqrt{4 \pi} r u(r)$, it is obvious
that $p\widetilde{U}(p)$ is a bounded and differentiable function for
all $p \geq 0$, and vanishes at $p = \infty$ by the Riemann-Lebesgue
lemma.$^6$ Everything is then quite meaningful in the integral in
(\ref{8e})~: there is absolute convergence at $p = \infty$, and the
principal value part is well-defined since $\widetilde{U}(p)$ is
differentiable. One can then show that if $k_{\nu} \to \infty$, the
principal value integral vanishes.$^{2,3}$. It follows that positive
energy bound states cannot go to infinity, and therefore, that they are
finite in number. In fact, this integral can also be written as a nice
Fourier cosine integral, as shown in Appendix A, if one assumes also $U
\in L^1(0, \infty )$~:

\beq \label{10e} \left \{ \begin{array}{l}  \varepsilon + 
\displaystyle{{2 \over \pi}}\ P \displaystyle{\int_0^{\infty}} \cdots =
\varepsilon + \displaystyle{\int_0^{\infty}} \omega (r) \cos kr\ dr\ ,
\\ \\ \omega (r) \in L^1(0 , \infty )\ . \end{array} \right . \eeq

This formula shows that, under the integrability condition on $U(r)$,
given in (\ref{7e}), the principal value integral is a bounded and
continuous function of $k$, and vanishes at infinity, so that the whole
expression goes to $\varepsilon (= \pm 1)$. Therefore, from (\ref{8e}),
there cannot be positive energy bound states beyond some value of $k^2$.
\par

In any case, as was shown by Gourdin and Martin,$^2$ one may have any
number of positive energy bound states by choosing $u(r)$ appropriately
through the inverse problem techniques for separable potentials.\par

Let us look now to the case where a local positive potential is also
present~:

\beq 
\label{11e}  
\left \{ \begin{array}{l}  \psi '' + k^2 \psi = V(r) \ \psi (k, r) + \varepsilon\ U(r) \displaystyle{\int_0^{\infty}} U(r') \ \psi (k, r') dr'\ .   \\ \\ 
\varepsilon = \pm 1\ , \displaystyle{\int_0^{\infty}} r\ V(r) \ dr < \infty\ , \quad V(r) \geq 0\ . \end{array} \right .
\eeq
\vskip 5 truemm

\noi Here, we assume that the Schr\"odinger equation with only the local potential~:

\beq 
\label{12e}  
\left \{ \begin{array}{l} \varphi '' + k^2\varphi = V \varphi \ , \\ \\
\varphi (k, 0) = 0 \ , \quad \varphi ' (k,0) = 1\ ,
 \end{array} \right .
\eeq
\vskip 5 truemm

\noi can be  solved explicitely and we know $\varphi (k, r)$. When $V =
0$, we get, of course, $\varphi = \sin kr/k$. Since $V$ is assumed
positive, there are no negative energy bound states, and one can show
that the set $\{\varphi (k, r); k \in [0, \infty )\}$ is complete in
$L^2(0, \infty )$,$^{1,3,4}$ and can be used to define integral
transforms quite similar to Fourier sine transform. Like $\sin kr/k$,
$\varphi (k, r)$ is an even entire function of $k$ of exponential type
$r$. In fact, we have, for every fixed $r > 0$,

$$\varphi (k, r)  \ \mathrel{\mathop =_{|k| \to \infty }}\ {\sin kr \over k} \left [ 1 + o(1)\right ] \ . \eqno({\rm 12.a})$$

\vskip 5 truemm
Also, it can be shown that, like for $\sin kr/k$, the zeros of
$\varphi$, for every fixed $r$, are all real if $V(r) > 0$, and,
therefore, because of (12.a) are given asymptotically by $k_n =
\pm n\pi/r$. One well-known example is, naturally, $V(r) = l (l+1)/r^2$, $l
\geq 0$, which leads to Hankel transform in which, instead of $\sin
kr$, one has to deal with the appropriate Bessel function. The
potential here is outside the class defined in (\ref{11e}), but one can
still show the completeness, as is well-known (E. C. Titchmarsh,
Eigenfunction Expansions I, Oxford University Press, 2nd ed., 1962. In
this book, one finds many examples of eigenfunction expansions related
to various differential equations of second order.). We can
define now

\beq \label{13e} \widetilde{U}(k) = \int_0^{\infty} U(r)\ \varphi (k,
r) dr \ . \eeq

\noi When $V = 0$, we go back, of course, to (\ref{9e}). Using now this
integral transform with (\ref{11e}), one gets then, that now the
positive energy bound states of (\ref{11e}) are given by the
simultaneous roots of the following two equations~:

\beq 
\label{14e}  
\left \{ \begin{array}{l}  \widetilde{U}(k_{\nu}) = 0\ , \\ \\ \varepsilon + \displaystyle{{2 \over \pi}} \ P \displaystyle{\int_0^{\infty}} \displaystyle{{\widetilde{U}^2 (p) \over p^2 - k_{\nu}^2}}\ \displaystyle{{p^2 \over |F(p)|^2}} \ dp = 0 \ , \end{array} \right .
\eeq

\vskip 5 truemm
\noi where $F(k)$ is the Jost function of the local potential $V$, i.e.
of the equation (\ref{12e}).$^{1,3}$ It is known that $F(k)$, which is
a continuous function for $k \geq 0$, never vanishes for $k \in [0,
\infty )$, and $F(\infty ) = 1$. Again, it is easily shown, under the
conditions (\ref{7e}) and (\ref{4e}) on $U(r)$ and $V(r)$, that the
principal value integral is well-defined.$^3$ The similarity with (\ref{8e}) is to be noticed here. So, one may hope that if a sufficient
condition on $U(r)$ is found to forbid positive energy bound states in
(\ref{7e}), a similar condition may be expected for (\ref{11e}), for a given $V(r)$. \\

\noi {\bf II. Absence of Positive Energy Bound States}\\

We consider first the simple case where $V = 0$, and so we have
(\ref{8e}) and (\ref{9e}). Now, a very simple condition to forbid
simultaneous roots of the two equations in (\ref{8e}) is to see whether
one can choose $U(r)$ in such a way as to have $\widetilde{U}(p) > 0$
for all $p \geq 0$. In this case, there cannot be any real common
roots. Here, one can use the following theorems for Fourier sine and cosine
transforms~:$^7$\\

\noi \underline{\bf Theorem 1 (Titchmarsh).} {\it Let $f(x)$ be non increasing over
$(0, \infty )$, integrable over $(0,1)$, and let $f(x) \to 0$ as $x \to
\infty$. Then $F_s (k) \geq 0$, where $F_s$ is the Fourier sine
transform of $f$. In fact, $F_s(k) > 0$ for $k > 0$ if $f(x)$ is strictly
decreasing.}\\

\noi \underline{\bf Theorem 2 (Titchmarsh).} {\it Let $f(x)$ be a bounded function,
which decreases steadily to zero as $x \to \infty$ and is convex. Then
$F_c(k)$, the Fourier cosine transform of $f(x)$, is positive and
belongs to $L^1(0, \infty )$.}\\

\noi The first  theorem applies directly to (\ref{8e}) and (\ref{9e}), and leads to the somewhat trivial~:\\

\noi \underline{\bf Theorem A.} {\it If $U(r)$ is a strictly decreasing
function of $r$, is $L^1(0, 1)$, and $rU(r) \in L^1(1, \infty )$, the
Schr\"odinger equation (\ref{7e}) has no positive energy bound states.
Note here that if $U$ is $L^1$ at the origin, $rU$ also is $L^1$, so
that the integrability condition on $rU$, shown in (\ref{7e}), reduces
to $rU \in L^1$ at infinity. And since $U$ is decreasing, one has, of course, $U(\infty ) = 0$.}\\

\noi \underline{\bf Remark 1.} In fact, any function $U(r)= ru(r)$ of positive
type satisfying the integrability conditions given in (\ref{4e}) would
lead, of course, to the same result. The Theorem 1 is just a
simple criterion. The purpose of Theorem A is to prepare
for what follows. \\

We have now to look at (\ref{11e}) to (\ref{14e}), where $V(r) > 0$. In
order to find a simple condition as above, we must first generalize
the theorem 1 of Titchmarsch to the integral transform (\ref{13e}).
A simple generalization is~:\\

\noi \underline{\bf Theorem 3.} {\it In order for $\widetilde{U}(k)$ defined by
(\ref{13e}) to be positive, it is sufficient that $U(r)$ be of the form

\beq \label{15e}  
U(r) = \varphi_0 (r) \int_r^{\infty} \varphi_0 (t) \ g(t) \ dt \int_r^t {du \over \varphi^2_0 (u)} \ ,
\eeq

\vskip 5 truemm
\noi where $\varphi_0 (r) = \varphi (k= 0, r)$, and $g(r)$ is any
positive function, which is such that $r^2g(r) \in L^1(0, 1)$, and
$rg(r) \in L^1(1, \infty )$. Moreover, from the assumptions on $g(r)$,
one gets also that $U(r) \in L^1(0,1)$, $rU(r) \in L^1(1, \infty )$,
and $U(r)$ is a decreasing function, so that $U(\infty ) = 0$.} \\

\noi The proof of this theorem will be given in Appendix B. \par

In order to make (\ref{15e}) more precise, we must of course show that
the last integral in the right-hand side is meaningful, i.e. $\varphi_0
(r) \not= 0$ for all $r > 0$, and that the whole integral is convergent
at $t = \infty$. These follow from the differential equation for
$\varphi_0(r)$, which is (\ref{12e}) at $k = 0$~:

\beq \label{16e} \left \{ \begin{array}{l} \varphi_0'' (r) = V(r) \
\varphi (r) \quad , \qquad V(r) > 0 \ ,\\ \\ \varphi_0 (0) = 0 \quad ,
\qquad \varphi '_0 (0) = 1 \ ,\end{array}\right .\eeq

\vskip 5 truemm
\noi and where one assumes that $rV(r) \in L^1(0, \infty )$. On the
basis of this assumption, one can show that$^{1,3}$

$$\varphi_0 (r) > 0\quad , \qquad \forall\ r > 0\ , \eqno({\rm 17.a})$$

\noi and 

$$\left \{ \begin{array}{l} \varphi_0 (r) = r[1 + o(1)] \ ,
\qquad \hbox{as $r \to 0$} \ ,\\ \\ \varphi_0 (r) = Ar+B+o(1) \ ,
\qquad \hbox{as $r \to \infty$} \ ,\end{array}\right .\eqno({\rm
17.b})$$
\vskip 5 truemm

\noi where $A > 1$, and $B < 0$. In short, $\varphi_0(r)$ is an
increasing convex function of $r$ since, from (17.a), $\varphi_0''
(r) > 0$, and it grows linearly as $r \to \infty$. Using the above
properties of $\varphi_0(r)$, it is now quite easy to show that the
right-hand side of (\ref{15e}) is quite meaningful under the conditions
given on $g(r)$ (Appendix B). \par

We introduce now the second independent solution of (\ref{16e})

$$\left \{ \begin{array}{l} \chi_0(r) = \varphi_0 (r) \displaystyle{\int_r^{\infty}
{du \over \varphi_0^2 (u)}}\quad , \qquad \chi_0(0) = 1\ , \\ \\
W(\varphi_0, \chi_0) = \varphi'_0 \chi_0 - \varphi_0 \chi '_0 = 1 \ .
\end{array} \right . \eqno(18)$$

\vskip 5 truemm

\noi From its definition, $\chi_0(r) > 0$ for all $r \geq 0$. Also,
since $\chi_0'' = V\chi_0$, $\chi_0(r)$ is, like $\varphi_0$, a convex
function of $r$. From the second part of (17.b), it is now easily seen
that

$$\chi_0 (\infty ) = {1 \over A} < 1\ . \eqno(19)$$

\vskip 5 truemm

\noi Since $\chi_0(0) = 1$, it follows that $\chi_0(r)$ is a decreasing
convex function of $r$. At any rate, using the definition of $\chi_0$
given in (18) in formula (\ref{15e}), we find

$$U(r) = \int_r^{\infty} \left [ \chi_0 (r) \ \varphi_0(t) -
\varphi_0(r) \ \chi_0 (t) \right ] g(t) \ dt \ . \eqno(20)$$

\vskip 5 truemm

\noi From this formula, it is immediately found that, under the assumptions of Theorem~3 on $g(r)$, one has (Appendix B)~:

$$U \in L^1(1,0)\quad , \qquad \hbox{and $U(\infty ) = 0$} \ . \eqno(21)$$

\noi Now, from (\ref{15e}), we have $U(r) > 0$, and differentiating
(20) twice, we find (remember that $g(r)$ is positive)

$$U''(r) - V(r) \ U(r) = g(r) \ , \eqno(22)$$

\noi which shows that $U(r)$ is also a convex function of $r$, and
since $U(0) > 0$, and $U(\infty ) = 0$, $U$ is a decreasing
convex function of $r$.\\

\noi \underline{\bf Remark 2.} As we see from the above analysis, $U(r)$ given by (\ref{15e}) is less
general than $U(r)$ of Theorem 1 of Titchmarsh where $U(r)$ had to be
only $L^1$ at $r = 0$, whereas here $U(0)$ is finite. Also, $U(r)$ of
Titchmarsh was only a decreasing function, whereas here we have our
$U(r)$ is even convex. The reason for all these shortcomings is that
Theorem 3 is a restricted form of the more general theorem whose proof
will be given in a separate paper. In any case, Theorem 3 now applies directly to (\ref{13e}), and leads to~:\\

\noi \underline{\bf Theorem B.} {\it Given $V(r)$, in order for
(\ref{11e}) to have no positive energy bound states, it is sufficient
for $U(r)$ to be of the form (\ref{15e}), where $g(r)$ is any positive
function such that $r^2g(r) \in L^1(0, 1)$ and $rg(r) \in L^1(1, \infty )$.}\\

\noi \underline{\bf Remark 3.} Condition (\ref{11e}) on $V(r)$ is sufficient, but
is not necessary in general. Examples for (\ref{13e}) are many. We just
mention the Hankel transform,$^7$ using Bessel functions $\sqrt{r}\
J_{\ell + {1 \over 2}}(r)$ instead of sine, which correspond to 

$$V(r) = {\ell (\ell + 1) \over r^2}\ , \quad \ell \geq 0 \ . \eqno(23)$$

\noi We have

$$\widetilde{F}_{\nu}(k) = \int_0^{\infty} f(r) \ \sqrt{kr}\ J_{\nu}(kr) dr \ ,\eqno(24)$$

\vskip 3 truemm
\noi where $\nu = \ell + {1 \over 2}$. Then, for $a > 0$, we have~:$^8$

$$f(r) = r^{-1/2}\quad , \quad \widetilde{F}_{\nu}(k) = k^{-1/2}\ ;$$

$$f(r) = r^{-1/2} \left ( r^2 + a^2\right )^{-1/2} \quad , \quad \widetilde{F}_{\nu} (k) = \sqrt{k}\ I_{{\nu \over 2}} \left ({1 \over 2} ak\right )K_{{\nu \over 2}} \left ( {1 \over 2} ka\right )\ ;$$

$$f(r) = r^{-1/2} \ e^{-ax}\quad , \quad \widetilde{F}(k) = {k^{{1 \over 2} - \nu} \over \left ( a^2 + y^2\right )^{1/2}} \left [ \left ( a^2 + y^2\right )^{1/2} - a \right ]^{\nu}\ ;$$

\noi and

$$f(r) = r^{-1/2} \ e^{-ax^2}\quad , \quad \widetilde{F}(k) = {\sqrt{\pi} \over 2}\ \sqrt{{k \over a}} \exp \left ( -{k^2 \over 8a}\right )  I_{{\nu \over 2}} \left ( {k^2 \over 8a}\right ) \ ;$$

\noi etc. It is known that $I_{\nu}$ and $K_{\nu}$ do not vanish on the positive real axis for $\nu > 0$.$^9$\\

\noi \underline{\bf Generalizations.}

 1. Using the results obtained in the papers of Mills and Reading,$^3$ it
is possible to generalize (\ref{15e}) to the case of a local potential
plus a finite sum of separable potentials. However, the conditions one
obtains are cumbersome, and we shall not reproduce them here.\\

2. So far, we have restricted ouselves to $\ell = 0$ ($S$-wave) in
(\ref{6e}). One can consider the case $\ell \not= 0$ along similar
lines, and one gets results similar to Theorems A and~B. Details will
be given in a separate paper.\\

3. In this paper, we have considered the case where $g(t)$ is a
function. However, $g(t)$ may be a generalized function. This will be
dealt with in details in the separate paper mentioned above. To
conclude, consider just the simplest case where $g(t) = \lambda \delta
(t - r_0)$, $\lambda$ and $r_0$ both positive. One finds then~:

$$U(r) = \lambda \varphi_0 (r_0) \varphi_0(r) \int_r^{r_0} {du \over
\varphi_0^2 (u)} \ \theta (r_0 - r) = \lambda \varphi_0 (r_0) \left [
\chi_0 (r) - \chi_0 (r_0) \right ] \theta (r_0- r) \ ,\eqno(25)$$

\noi which has a finite range, and is finite at the origin since
$\chi_0 (1) = 1$. For $g(t)$ given by a finite sum of delta functions,
one gets a finite sum of such $U(r)$. \\

\noi {\large \bf Acknowledgments} \par \vskip 0.5 truecm

One of the authors (KC) would like to thank the Department of
Mathematics of the Tokyo University of Science, and Professor Kenro
Furutani, for their warm hospitality, and financial support.

 \newpage
\noi {\bf Appendix A.}\\

We have, from our assumptions, that $U(r)$ belongs to the
following class~: 

$$U(r) \in L^1(0, \infty )\quad ; \quad rU(r) \in L^1(0, \infty ) \ . \eqno({\rm A.1})$$

\noi Let us assume first that $U(r) > 0$, and define

$$\widetilde{u}(p) \equiv p\ \widetilde{U}(p) = \int_0^{\infty} U(r) \sin pr\ dr \ . \eqno({\rm A.2})$$

\noi The function $\widetilde{u}(p)$ is a bounded continuous function,
and $\widetilde{u}(0) = \widetilde{u}(\infty ) = 0$. But we have more.
Indeed, differentiating (A.2) with respect to $p$ under the integral
sign, we get

$$\dot{\widetilde{u}}(p) = \int_0^{\infty} [r\ U(r)] \cos pr\ dr \ , \eqno({\rm A.3})$$

\noi which is again a bounded continuous function, and
$\dot{\widetilde{u}}(\infty ) = 0$. Therefore, $\widetilde{u}(p)$ is,
in fact, $C^1[0, \infty )$. If we introduce now

$$W(r) = \int_r^{\infty} U(t)\ dt \ , \eqno({\rm A.4})$$

\noi because of second part of (A.1), it is immediately seen that, $W$ is a
bounded and continuous function for $r > 0$, and

$$W(r) \in L^1(0, \infty ) \ , \qquad \lim_{r\to 0, \infty} \ r \ W(r) = 0 \ . \eqno({\rm A.5})$$

\noi Introducing $U = - W'$ in (A.2), and integrating by parts, we find that $\widetilde{u}(p)$ can also be written as 

$$\widetilde{u}(p) = p \int_0^{\infty} W(r) \cos pr \ dr \ . \eqno({\rm A. 6})$$

\noi Now, if we assume, to begin with, that $U(r)$ is also a decreasing function, it follows that $W(r)$ is bounded and convex, and
$W(\infty ) = 0$. Therefore, from Theorem 2, we have

$${\widetilde{u}(p) \over p} \in L^1(0, \infty )\ , \eqno({\rm A.7})$$

\noi which shows that the integral in (\ref{8e}) is absolutely
convergent at $p = \infty$. Moreover, since $\widetilde{u}(p)$ is
$C^1$, there is also no problem for the existence and even H\"older
continuity of the principal-value integral.$^6$\par

Let us now write the integral in (\ref{8e}) as follows~:

$$G(k) = P \int_0^{\infty} {\widetilde{u}^2(p) \over p^2 - k^2} dp = P \int_0^{\infty} {\widetilde{u}^2(p) \over 2p} \left [ {1 \over p - k} + {1 \over p + k} \right ] dp \ . \eqno({\rm A.8})$$

\noi Changing $p$ to $-p$ in the second integral, it is easily found that 

$$G(k) = {1 \over 2} P \int_{-\infty}^{\infty} {\widetilde{u}(p) \over p}\ {\widetilde{u}(p) \over p - k} \ dp \ . \eqno({\rm A.9})$$

\noi If we use now (A.2) for one $\widetilde{u}(p)$, and (A.6) for the second one, we find

$$G(k) = {1 \over 2} \int_0^{\infty} U(r) \ dr \int_0^{\infty} W(r') \ dr' \ P \int_{-\infty}^{\infty} {\sin pr\ \cos pr' \over p - k}\ dp \ . \eqno({\rm A.10})$$

\noi The change of the order of integrations is justified because both
$U(r)$ and $W(r)$ are $L^1(0, \infty )$.$^6$ Now, since 

$$\sin pr \cos pr' = {1 \over 2} \left [ \sin p(r+r') + \sin p(r-r')\right ] \ , \eqno({\rm A.11})$$

\noi and$^6$

$$P \int_{-\infty}^{\infty} {\sin xy \over y - y_0} \ dy = \pi \cos x y_0 \ . \eqno({\rm A.12})$$

\noi We finally have

$$G(k) = {\pi  \over 2} \int_0^{\infty} U(r) \cos kr\ dr \int_0^{\infty} W(r') \cos kr'\ dr' = {\pi  \over 2} \widetilde{U}_c(k)\ \widetilde{W}_c(k) \ . \eqno({\rm A.13})$$

\noi The first integral being a bounded and continuous function, and
the second one $L^1(0, \infty)$ by the second theorem of Titchmarsh, as
we saw before, it follows that $G(k)$ is also $L^1(0, \infty )$. We can
now write $G(k)$ as a Fourier cosine transform$^{6,10}$

$$G(k) = \int_0^{\infty} \omega (r) \cos kr\ dr \ , \eqno({\rm A.14})$$

\noi where $\omega (r)$ is given by the convolution

$$\omega (r) = {\pi  \over 2} \int_0^{\infty} U(t) \left [ W(|r - t|) + W(r+t) \right ] dt\ . \eqno({\rm A.15})$$

\vskip 5 truemm

\noi And since both $U$ and $W$ are $L^1(0, \infty)$, it follows, from
the convolution theorem of two $L^1$ functions, that $\omega (r)$ is
also $L^1(0, \infty )$.$^6$\par

So far, we have been assuming that $U'(r) < 0$. However, in (A.15), no
reference is made to the derivative of $U$. If $U(r)$ satisfies only
(A.1), again all factors inside the integral in (A.15) are $L^1(0,
\infty )$, and so is $\omega (r)$. One may expect therefore that (A.14)
and (A.15) are true under (A.1) only. The direct proof of this
assertion needs more elaborate reasoning by using the methods of
ref.$^6$, chapter 2. We shall not reproduce it here. \par

We come now to the sign of $U(r)$ itself. So far, we have been assuming
$U(r)$ to be positive. However, in the main text, the assumption we
made is only the one shown in (\ref{7e}), with no reference to the sign
of $U(r)$. We have therefore to extend our result to the case where
$U(r)$ is oscillating. But this is all easy. Indeed, we can separate
the positive and negative parts of $U(r)$, and write

$$U(r) = U_+(r) - U_-(r)  \eqno({\rm A.16})$$

\noi where both $U_+$ and $U_-$ are positive, and, of course, satisfy
separately (\ref{7e}). We have now

$$\widetilde{U}^2(p) = \widetilde{U}^2_+(p) + \widetilde{U}^2_-(p) - 2
\widetilde{U}_+(p) \widetilde{U}_-(p) \ . \eqno({\rm A.17})$$

\noi It is now trivial to apply our previous reasoning separately to
each of the three terms here. To summarize, introducing $W_{\pm}$ as in
(A.4), and reducing our assumptions (A.1) to their essential parts, we
have \\

\noi \underline{\bf Theorem 4.} {\it Under the assumptions

$$U(r) \in L^1(0, 1) \ , \quad rU(r) \in L^1(1, \infty )\ , \eqno({\rm A.18})$$

\noi the second formula in (\ref{8e}) can be written as

$$\varepsilon + {2 \over \pi} P \int \cdots = \varepsilon + {\pi \over 2} \int_0^{\infty} \omega (r) \cos kt \ dt \ , \eqno({\rm A.19})$$

\noi where

$$\omega (r) = {\pi \over 2} \int_0^{\infty} U_+ (t) \left [ W_+ (|r-t|) + W_+ (r+t)\right ] dt $$
$$+ {\pi \over 2} \int_0^{\infty} U_-(t) \left [W_- (|r-t|) + W_- (r + t) \right ] dt$$
$$- \pi \int_0^{\infty} U_-(t) \left [ W_+ (|r -t|) + W_+ (r+t) \right ) dt \ , \eqno({\rm A.20})$$

\vskip 5 truemm
\noi and $\omega (r) \in L^1(0, \infty)$. In the third integral, one
can, of course, exchange $U$ and $W$, and have $U_+(t) [W_- \cdots
]$.}

\newpage
\noi {\bf Appendix B.}\\

\noi \underline{\bf Proof of Theorem 3.} We wish to show that given
$V(r) > 0$, which defines the integral transform (\ref{13e}), and under
suitable conditions on $U(r) > 0$, $\widetilde{U}(k)$ is positive. For
this purpose, we use the following integral representation for $\varphi
(k, r)$, which comes from the Gel'fand-Levitan theory of inverse
problems~:$^{1,3}$

$$\varphi (k,r) = {\sin kr \over k} + \int_0^{r} K(r, x) {\sin kx \over
k}\ dx \ . \eqno({\rm B.1})$$

\vskip 5 truemm

\noi The kernel $K(r, x)$, defined only for $0 \leq x \leq r$,
satisfies the Volterra integral equation~:$^{11}$

$$K(r, x) = {1 \over 2} \int_{{r-x \over 2}}^{{r+x \over 2}} V(s) \ ds
+ \int_{{r-x \over 2}}^{{r+x \over 2}} ds \int_0^{{r-x\over 2}} V(s +
u) \ K (s+u, s-u)\ du \ .\eqno({\rm B.2})$$

It can be shown that this Volterra integral equation can always be
solved by iteration, and lead to an absolutely convergent series,
provided that $rV(r) \in L^1(0 , \infty )$.$^{1,3}$ Moreover, one gets
the upper bound (remember that here $V(r)$ is positive)~:

$$K(r,x) \leq {1 \over 2} \int_{{r-x \over 2}}^{{r + x \over 2}} V(s)
\exp \left [ \int_0^{{r+x \over 2}} u V(u) du \right ] ds \leq {1 \over
2} \ e^{\int_0^{\infty} u V(u) du} \int_{{r-x \over 2}}^{{r + x \over
2}} V(s) ds\ . \eqno({\rm B.3})$$

We can use now (B.1) in the right-hand side of (\ref{13e}), and we get,
with a slight change of notations

$$\left \{ \begin{array}{l} \widetilde{U}(k) = \displaystyle{\int_0^{\infty}} f(x)
\displaystyle{{\sin kx \over k}}\ dx \ , \\ \\ f(x) = U(x) + \displaystyle{\int_x^{\infty}} K(r, x)\
U(r)\ dr \ . \end{array} \right . \eqno({\rm B.4})$$

\noi The exchange of the order of integrations in going from
(\ref{13e}) to (B.4) is legitimate because of the bound (B.3), which
shows that $K(\infty ,x) = 0$ for all $x > 0$, and by the assumption
that $U(\infty ) = 0$, so that the integrals are all absolutely
convergent at the upper limit.\par

Now, in (B.4), $\widetilde{U}(k)$ is a Fourier sine transform, and we
wish therefore to apply the Theorem 1 of Titchmarsh. We have therefore
to show that $f(x)$ satisfies all the requirements of that theorem,
namely~:

$$\left \{ \begin{array}{l} f(x) > 0 \ , \\ f(x) \in L^1(0,1)\ , \\ f(x) \ \hbox{is steadily decreasing, and}\\ f(\infty ) = 0 \ . \end{array} \right . \eqno({\rm B.5})$$

Now, since $K(r, x)$ is positive in (B.4), in order to secure that $f(x)$ is also positive, it is sufficient to assume that

$$U(r) > 0 \ . \eqno({\rm B.6})$$

Let us now check the second statement in (B.5). In (B.4), both $U$ and
$K$ being positive, it is obvious that to secure that $f(x) \in
L^1(0,1)$, we must assume

$$U(r) \in L^1(0,1) \ .\eqno({\rm B.7})$$

\noi That the integral $\int_x^{\infty} KU dr$ in (B.4) is also $L^1(0,1)$ follows now from the positivity of $K$ and $U$, (B.7) and

$$K(r,0) = 0  \qquad \hbox{uniformly in $r$}\ , \eqno({\rm B.8})$$

\noi which is an obvious consequence of (B.2) and (B.3).\par

Concerning the last statement in (B.5), $f(\infty ) = 0$, it is obvious
on (B.4) and the fact that both $U$ and $K$ are positive, that we must
assume

$$U(\infty ) = 0 \ , \eqno({\rm B.9})$$

\noi and this is sufficient. \par

It remains to show that $f(x)$ is a decreasing function. From (B.4), we have~:

$$f'(x) = U'(x) - K(x, x)\ U(x) + \int_x^{\infty} {\partial K (r, x) \over \partial x}\ U(r)\ dr \ . \eqno({\rm B.10})$$

\noi Now, differentiating (B.2) with respect to $r$ and $x$, we also find

$$\left \{ \begin{array}{l} \displaystyle{{\partial K(r, x) \over
\partial r}} + \displaystyle{{\partial K(r, x) \over \partial x}} =
F(r, x) \ , \\ \hfill ({\rm B.11}) \\ F(r,x) = \displaystyle{{1 \over 2}} V\left (
\displaystyle{{r+ x \over 2}} \right ) + \int_0^{{r-x \over 2}} V\left
( \displaystyle{{r+x \over 2}} + u \right ) K \left ( \displaystyle{{r+
x \over 2}} + u, \displaystyle{{r+ x \over 2}} - u\right ) du> 0\
.\end{array} \right .$$

\noi Extracting $\partial K/\partial x$, using it in (B.10), and
integrating by parts with respect to $r$, we finally find that we must
have

$$f'(x) = U'(x) + \int_x^{\infty} K(r, x) U'(r) dr + \int_x^{\infty}
F(r,x)\ U(r)\ dt < 0\ . \eqno({\rm B.12})$$

This condition is, obviously, very complicated, and nothing simple on
$U$ can be easily obtained from it. Let us therefore assume that $U''$
exists also. Differentiating (B.12), and using (B.11) at $r = x$~:

$$\left [ {\partial K(r, x) \over \partial r} + {\partial K(r, x) \over
\partial x}\right ]_{x=r} = F(r, r) = {1 \over 2} V(r) \ , \eqno({\rm
B.13})$$

\noi we find

$$f''(x) = U''(x) - {1 \over 2} V(x) \ U(x) - K(x, x)\ U'(x) -
{\partial K \over \partial x}(x,x) \ U(x) + \int_x^{\infty} {\partial^2
K(r, x) \over \partial x^2} \ U(r)\ dr \ . \eqno({\rm B.14})$$

It is now well-known that $K(r,x)$ satisfies the differential
equation$^{1,3}$

$${\partial^2 K(r, x) \over \partial r^2} - {\partial^2 K(r,x) \over
\partial x^2} = V(r)\ K(r,x) \ , \eqno({\rm B.15})$$

\noi which can also be obtained from (B.2). If we replace now
$\partial^2K/\partial x^2$ obtained from (B.15) in (B.14), integrate by
parts twice the integral containing $\partial^2K/\partial r^2$, and use
the fact that, because of (B.2), (B.3), and (B.9), all the integrated
terms vanish at $r = \infty$, we finally find

$$f''(x) = [U''(x) - V(x) \ U(x)] + \int_x^{\infty} K(r, x) [U''(r) -
UV]dr\ . \eqno({\rm B.16})$$

\noi It follows that, if we assume

$$U''(x) - V(x) \ U(x) > 0\ , \eqno({\rm B.17})$$

\noi then

$$f''(x) > 0 \ , \eqno({\rm B.18})$$

\noi that is, $f(x)$ is a convex function. Now, as we saw before,
$f(x)$ is positive, and $f(\infty ) = 0$. The convexity of $f(x)$
secures then that$^{12}$

$$f'(x) < 0 \ . \eqno({\rm B.19})$$

\noi We have therefore completed all the sufficient conditions to
secure (B.5), namely (B.6), (B.7), (B.9), and (B.17). We can therefore
apply the theorem 1 of Titchmarsh to $f(x)$, and we get~:\\

\noi \underline{\bf Theorem 5.} {\it Under the conditions (B.6), (B.7),
(B.9), and (B.17), we have}

$$\widetilde{U}(k) = \int_0^{\infty} f(r) {\sin kr \over k} \ dr > 0\ .
\eqno({\rm B.20})$$ \vskip 5 truemm

In order to prove Theorem 3 of the main text, we must now study more
(B.17), which we write as

$$U''(x) - V(x) \ U(x) = g(x) \quad , \qquad g(x) > 0 \ . \eqno({\rm
B.21})$$

\noi For the time being, $g(x)$ is, of course, arbitrary. However, it
must be such that the solution of (B.21) satisfies (B.6), (B.7), and
(B.9). Equation (B.21) being a simple inhomogeneous linear differential equation of
second order, it is well-known, and can be checked in a straightforward
manner, that a solution satisfying (B.9), i.e. $U(\infty ) = 0$, is
given by

$$U(r) = \int_r^{\infty} \left [ \chi_0 (r) \varphi_0 (t) - \varphi_0
(r) \chi_0 (t) \right ] g(t) \ dt = \varphi_0 (r) \int_r^{\infty}
\varphi_0(t) \ g(t) \ dt \int_r^t {du \over \varphi^2_0(u)}\ ,\eqno({\rm
B.22})$$

\noi where $\varphi_0$ and $\chi_0$ were defined in the main text. From
the properties of $\varphi_0$ and $\chi_0$ we established there, one
finds easily, from the last expression in (B.22), that

$$U(r) \ \mathrel{\mathop \simeq_{r \to \infty }}\ Ar \int_r^{\infty}
{1 \over A} \ g(t) {1 \over A^2} \left ( {1 \over r} - {1 \over t}\right
) dt \ . \eqno({\rm B.23})$$

\vskip 5 truemm

\noi Since $g(t) > 0$, it follows that, in order to have the
convergence of the integral at infinity, we must have $g(t) \in
L^1(\infty )$. And this secures, of course, that $U(\infty ) = 0$.
Similarly, if we wish to have $U(r) \in L^1$ at infinity, we must have
$tg(t) \in L^1$ at infinity. Indeed, taking $B$ large enough, we have
$\int_B^{\infty} U(r) dr \cong \int_B^{\infty}$[(B.23)]$dr$. Everything
being positive in (B.23), we can exchange now the orders of integration
in $r$ and $t$. Then it is immediately seen that if $tg(t)$ is $L^1$,
so is $U(r)$. For having $rU(r)\in L^1$ at infinity, (\ref{7e}), and needed
also in Appendix A, one must have $t^2g(t) \in L^1$ at infinity, etc.
\par

Let us now look at what happens at $r = 0$, for we have to secure
(B.7). We can use again now the behaviour of $\varphi_0$ and $\chi_0$,
given in the main text, in the middle expression in (B.22). One finds
easily, that

$$U(r) \ \mathrel{\mathop \simeq_{r \to 0 }}\left ( \int_{r\to
0}^{\infty} \varphi_0(t) \ g(t) \ dt\right )  - \left ( r \int_{r \to
0}^{\infty} \chi_0 (t)\ g(t) \ dt\right )   \ . \eqno({\rm B.24})$$

\vskip 5 truemm

\noi The first integral here is finite if $tg(t)$ is $L^1(0)$, and the
second one also because we can put $r$ inside the integral and make it
larger. To have only $U \in L^1$, at the origin, it is obvious first
that it is sufficient to replace in (B.24) the two integrals by
$\int_{r\to 0}^1 \cdots dt$. Then we have~:

$$\int_0^1 U(r) \ dr = \int_0^1 dr \left [ \int_r^1 [\varphi_0 (t) \ g(t) -
r \ \chi_0 (t) \ g(t) \right ] dt \ . \eqno({\rm B.25})$$

\vskip 5 truemm

\noi Again, all the functions here being positive, we can excahnge the
orders of integrations in each double integral. One finds then
immediately that $t^2g(t) \in L^1(0, 1) \Rightarrow U \in L^1(0, 1)$. In
short, we have

$$\left \{ \begin{array}{l} t^{\alpha} g(t) \in L^1(1, \infty )
\Rightarrow r^{\alpha - 1} \ U(r) \in L^1(1, \infty )\quad , \quad \alpha = 1, 2 \ ,\\ \\ t^2g(t) \in
L^1(1, 0) \Rightarrow U(r) \in L^1(1, 0)\ . \end{array} \right . \eqno
({\rm B.26})$$

\vskip 5 truemm
\noi This completes the proof of the properties of $g(t)$ in Theorem 3,
in Theorem B, and elsewhere, and provides sufficient conditions on the
properties of $U(r)$ needed in the main text, and in Appendix A .\\

\noi \underline{\bf Remark.} In all rigor, $K(x, x)$ in (B.5) is
infinite if $V(r)$ is not integrable at $r = 0$, as is seen on (B.2)
and (B.3). This may happen since we assume only $rV(r)$ to be
integrable at $r = 0$. However, one may first regularize the potential
at $r = 0$, and proceed as we did. Then, it can be seen that the final
form (B.7) is quite general, and independent of whether $V(x)$ is
integrable or not at the origin. One can therefore remove the
regularization in (B.7), and so condition (\ref{15e}) is quite general.

\newpage
\centerline{\large \bf References} \par \vskip 1 truecm

\begin{enumerate}
\item R. G. Newton, Scattering Theory of Waves and Particles
(Springer-Verlag, New York, 1982). We shall quote often this book. See
especially chapter 12. For a review paper on the use of separable
potentials in Faddeev equations, the interested reader can consult~: S.
Oryu, Phys. Rev. {\bf C27}, 2500 (1983), where full references to
earlier works can be found.

\item M. Gourdin and A. Martin, C. R. Acad. Sc. Paris, {\bf 244}, 1329
(1957)~; Nuovo Cimento {\bf 8}, 699-719 (1958). See also$^3$. In these
papers, the inverse problem for separable potentials, including the
possibility of positive energy bound states, was solved for the first
time, and the complete solution was given explicitely.

\item K. Chadan and P. C. Sabatier, Inverse Problems in Quantum
Scattering Theory, 2nd ed. (Springer-Verlag, Berlin, 1989),
chapter VIII, where full references to the original works of Y. Y.
Yamaguchi, M. Gourdin and A. Martin, K. Chadan, M. Bolsterli and
J. Mackenzie, F. Tahakin, R. L. Mills and J. F. Reading, etc, on separable potentials can be found. We shall
follow the notations of this book.

\item B. Simon, Quantum Mechanics for Hamiltonians defined as Quadratic
Forms (Princeton U. P., Princeton, NJ, 1971), pp. 89-93. See also A.
Galindo and P. Pascual, Quantum Mechanics, Vol. I (Springer, Berlin
1990), pp. 233-234.

\item F. Zirilli, Il Nuovo Cimento, {\bf 34A}, 385 (1976).

\item E. C. Titchmarsh, Introduction to the Theory of Fourier Integrals (Oxford U. P., 2nd ed., 1959).

\item Ref. 6, p. 169, Theorem 123, p. 170, Theorem 124.

\item Tables of Integral Transforms, Vol. II, A. Erd\'elyi, Editor (Mc Graw-Hill, New York, 1954). See pages 22, 23, 28 and 29.

\item Higher Transcendental Functions, Vol. II, A. Erd\'elyi, Editor (Mc Graw-Hill, New York, 1953).

\item I. N. Sneddon, Fourier Transforms (Mc Graw-Hill, New York 1951), pp. 23-25.

\item Ref. 3, pp. 43-45.

\item G. Valiron, Th\'eorie des fonctions, 2\`eme \'ed. (Masson et Cie, Paris 1948). H. L. Royden, Real Analysis, 2nd ed. (McMillan, New York, 1968).
\end{enumerate}
\end{document}